\documentclass[10pt,conference]{IEEEtran}

\usepackage{graphicx} 
\usepackage{todonotes}
\usepackage{listings}
\usepackage{multirow}
\usepackage{amsmath}
\usepackage{stmaryrd}
\usepackage{mdframed}
\usepackage{tcolorbox}
\usepackage{multicol}
\usepackage{caption}
\usepackage{xspace}
\usepackage{url}
\usepackage{amsthm}
\usepackage{dirtree}
\usepackage[bottom]{footmisc}

\lstdefinestyle{mystyle}{
	basicstyle=\ttfamily\scriptsize,
	breakatwhitespace=false,         
	breaklines=true,                 
	captionpos=b,                    
	keepspaces=true,                 
	numbers=left,                    
	numbersep=5pt,                  
	showspaces=false,                
	showstringspaces=false,
	showtabs=false,                  
	tabsize=2,
	frame=single,
	xleftmargin=3.0ex
}

\newcommand{\daleq}{\textit{daleq}\xspace}
\newcommand{\Daleq}{\textit{Daleq}\xspace}
\newcommand{\souffle}{\textit{souffle}\xspace}
\newcommand{\bfs}{\textit{obfs}\xspace}
\newcommand{\gaoss}{\textit{gaoss}\xspace}
\newcommand{\mvnc}{\textit{mvnc}\xspace}
\newcommand{\mvncvsbfs}{\textit{mvnc-vs-bfs}\xspace}
\newcommand{\mvncvsgaoss}{\textit{mvnc-vs-gaoss}\xspace}
\newcommand{\jnorm}{\textit{jnorm}\xspace}
\newcommand{\javap}{\textit{javap}\xspace}
\newcommand{\JavaBEPEnv}{\textit{JavaBEPEnv}\xspace}
\newcommand{\ASM}{\textit{ASM}\xspace}

\begin{document}
\title{DALEQ - Explainable Equivalence for Java Bytecode}

\author{
	 \IEEEauthorblockN{Jens Dietrich}
     \IEEEauthorblockA{Victoria University of Wellington
     \\Wellington, New Zealand
     \\ jens.dietrich@vuw.ac.nz}
  
	\and
	\IEEEauthorblockN{Behnaz Hassanshahi}
	\IEEEauthorblockA{\textit{Oracle Labs Australia} \\
	Brisbane, Australia \\
	behnaz.hassanshahi@oracle.com}
}

\maketitle
\thispagestyle{plain}
\pagestyle{plain}

\begin{abstract}
	
The security of software builds has attracted increased attention in recent years in response to incidents like \textit{solarwinds} and \textit{xz}. Now, several companies including Oracle and Google rebuild open source projects in a secure environment and publish the resulting binaries through dedicated repositories. This practice enables direct comparison between these rebuilt binaries and the original ones produced by developers and published in repositories such as Maven Central. These binaries are often not bitwise identical; however, in most cases, the differences can be attributed to variations in the build environment, and the binaries can still be considered equivalent. Establishing such equivalence, however, is a labor-intensive and error-prone process.

While there are some tools that can be used for this purpose, they all fall short of providing provenance, i.e. readable explanation of why two binaries are equivalent, or not.  To address this issue, we present \daleq, a tool that disassembles Java byte code into a relational database, and can normalise this database by applying datalog rules.  Those databases can then be used to infer equivalence between two classes. Notably,  equivalence statements are accompanied with datalog proofs recording the normalisation process. We demonstrate the impact of \daleq in an industrial context through a large-scale evaluation involving 2,714 pairs of jars, comprising 265,690 class pairs. In this evaluation, \daleq is compared to two existing bytecode transformation tools. Our findings reveal a significant reduction in the manual effort required to assess non-bitwise equivalent artifacts, which would otherwise demand intensive human inspection. Furthermore, the results show that \daleq outperforms existing tools by identifying more artifacts rebuilt from the same code as equivalent, even when no behavioral differences are present.
  
 
\end{abstract}

%
%


\section{Introduction}
\label{sec:introduction}

Software is the foundation of modern digital infrastructure. 
Modern software systems are assembled from existing components, using automated processes like continuous integration and deployment. 
This has created new security problems as both components and processes can inject vulnerabilities into systems. 
Examples include \textit{solarwinds}, \textit{codecov}, \textit{equifax} and \textit{log4shell}~\cite{ellison2010evaluating,martinez2021software,enck2022top} for compromised component, and \textit{xcodeghost}, \textit{ccleaner}, \textit{shadowpad}, \textit{shadowhammer} and \textit{xz}~\cite{martinez2021software,andreoli2023prevalence,xz,ohm2020backstabber} for compromised processes. 

Initiatives like reproducible builds~\cite{reproduciblebuild, lamb2021reproducible, enck2022top, butler2023business, fourne2023s} aim to enhance software supply chain security by rebuilding packages from source code independently and comparing the resulting binaries. The objective is to produce identical, bitwise equivalent binaries. Achieving this is not straightforward and often requires significant engineering effort~\cite{macho2021nature, shi2021experience, mukherjee2021fixing, ren2022automated, drexel2024reproducible, randrianaina2024options, keshani2024aroma}. Industry initiatives such as Google Assured Open Source (\gaoss) and Oracle Build-From-Source (\bfs), among others, also focus on rebuilding open-source artifacts from source on secure and hardened build services. While the goal is not necessarily to achieve bit-by-bit equality with the reference binaries built and released by the open-source developers (referred to as \textit{reference binaries}), substitutability remains a key requirement. I.e. a binary should be replaceable by the alternatively built binary without changing the behaviour of downstream programs. Substitutability is trivially achieved if the binaries are identical. If not, this becomes more complex, since behavioural equality is undecidable. 
However, it is still possible to under-approximate behavioural equality by devising an equivalence relation $\simeq$ between binaries (i.e. in Java) where $b_1 \simeq b_2$ implies that $b_1$ and $b_2$ have the same behaviour.

This idea was introduced in our previous work~\cite{dietrich2024levels}, where we evaluated how equivalences based on existing tools, such as decompilers, disassemblers, and bytecode normalization tools, performed across various datasets. In this paper, we build upon that work and present \daleq, a novel approach that not only establishes equivalence between Java binaries but also provides a provenance, i.e., an explanation of why two binaries are either equivalent or not. Conceptually, \daleq is similar to normalization tools like \jnorm~\cite{schott2024JNorm} and \JavaBEPEnv~\cite{xiong2022towards}, but with the added benefit of explainability features that help users gain confidence in the results.

The technique presented here has been successfully applied to evaluate open-source artifacts used and built from source in Oracle’s Graal Development Kit for Micronaut (GDK)~\footnote{https://www.oracle.com/developer/gdk-developers/}. This application underscores its practical impact on enhancing software supply chain security, addressing the critical need for product teams to verify that binaries remain uncompromised while maintaining functional equivalence.

This paper presents the following contributions:

\begin{enumerate}
	\item \daleq provides provenance information supporting both equivalence and non-equivalence statements that can be used by security engineers to assess, validate and trust its outputs.
	\item \daleq is based on datalog, following a widely used approach in static program analysis. The rule-based constructions assures its high correctness, i.e., while there is no formal proof of soundness, it is highly unlikely that \daleq will flag pairs of binaries with different behaviour as equivalent.   
	\item the comparative evaluation suggests that \daleq significantly outperforms the state-of-the-art tool \jnorm. This directly translates into significant cost savings of engineering time required to assess alternative build outputs. Since \JavaBEPEnv is not available to us, we have not been able to evaluate \daleq against this tool. 
\end{enumerate}

This paper is organised as follows. We briefly review related work including publications and tools in Section~\ref{sec:relatedwork}. This is followed by a detailed discussion of the design and implementation of \daleq in Sections~\ref{sec:design} and \ref{sec:implementation}, respectively. We then evaluate \daleq against two similar tools on a dataset consisting of jars from Maven Central, compared with jars built by \gaoss and \bfs in Section~\ref{sec:evaluation}. This is followed by a conclusion.

\section{Related Work}
\label{sec:relatedwork}

\subsection {Diff Tools}

\textit{Diffoscope}~\footnote{\url{https://diffoscope.org/}} is a general-purpose diff tool that can also be applied to (the content) of jar files. It directly uses bytecode, without applying abstractions. This makes it sensitive to even minor changes,  and creates noise, e.g. it reports changes caused by platform specific new line separators in metadata files. It also reports file attributes. In contrast, \daleq focuses on differences that influence program behaviour, and on minimising noise (i.e. differences that do not influence program behaviour).

\textit{JarDiff}~\footnote{\url{https://github.com/lightbend-labs/jardiff/}} is a specialized tool designed to show differences in jar files built from Java programs . It shares some conceptual similarities with \daleq, as both tools generate differences in bytecode abstractions through transformations, specifically using \textit{scalap} and the \textit{asm textifier}.  However, \daleq goes a step further by offering a more comprehensive approach, including detailed equivalence checks and advanced normalization techniques.
While \textit{JarDiff} focuses on establishing some equivalences (e.g., resolving constant pool ordering issues) without applying normalization, \daleq's equivalence analysis involves additional steps, including the application of rules after extracting the EDBs. This makes \daleq a more robust solution for analyzing and comparing Java binaries.

There are several specialised diff tools available to check for changes in jars corresponding to different versions of the same program, including \textit{revapi}~\footnote{\url{https://revapi.org/}}, \textit{japicmp}~\footnote{\url{https://siom79.github.io/japicmp/}}, and \textit{clirr}~\footnote{\url{https://github.com/ebourg/clirr}}. Those tools focus on detecting API changes violating Java source and binary compatibility rules that can lead to compile- and link-time errors in downstream clients~\cite{dietrich2014broken,raemaekers2014semantic,jayasuriya2023breaking}. Those tools are not suitable to detect subtle semantic changes that may indicate a compromised. In general, when assessing artifacts rebuilt from the same source code, the APIs rarely change.

\subsection{Binary Equivalence Levels}

In our prior work~\cite{dietrich2024levels}, we addressed the strict definition of reproducible builds, which primarily focuses on bitwise equivalence, by proposing a more practical set of levels for establishing binary equivalence. Building on this conceptual foundation, we introduce a technique for achieving level 3 equivalence in this study. Our previous work also included a large-scale evaluation of existing tools, such as decompilers, disassemblers, and \jnorm, that can be adapted for equivalence checking. Among these, \jnorm demonstrated the strongest performance and continues to be a key component in our approach. In this paper, we extend the previous analysis by incorporating a discussion of potential spurious equivalences, which can arise during the equivalence checking process. The equivalence levels defined in~\cite{dietrich2024levels} form the basis of the methodology applied in this work.

Building further on this line of research, Dietrich et al.~\cite{dietrich2025towards} explored the synthesis of witnesses for non-equivalence statements through automated test case generation.

\subsection{Binary Normalization and Transformation Techniques}

Schott et al. introduce \jnorm~\cite{schott2024JNorm}, a tool designed to normalize Java bytecode using the \textit{Jimple} intermediate representation provided by Soot~\cite{vallee2010soot}. This approach exemplifies a transformation function that aids in the construction of equivalence relations for binary comparisons. We adopt and evaluate this methodology in the context of binary equivalence for software supply chain security. It is important to note that \jnorm was originally developed for \textit{code similarity analysis}. As such, it applies abstractions that simplify certain details, potentially influencing the semantics of the program. We will explore examples of these abstractions and their impact on binary equivalence in the following sections.

Xiong et al. investigate barriers to the reproducibility of Java-based systems~\cite{xiong2022towards}. They analyse sources of bytecode variability across builds and propose \textit{JavaBEPFix}, a tool designed to eliminate certain build-induced differences. In contrast to \jnorm and disassembler- or decompiler-based transformations, \textit{JavaBEPFix} produces normalised bytecode rather than an abstract representation. Nonetheless, it aligns with the framework of transformation-based equivalence~\cite{dietrich2024levels}. Unfortunately, we were unable to include \textit{JavaBEPFix} in our evaluation, as the tool is not publicly available.

A substantial body of work addresses bytecode similarity~\cite{baker1998deducing,ji2008plagiarism,davis2010whence,keivanloo2014sebyte,chen2014achieving,yu2019detecting,schneider2022experimental}. However, similarity-based approaches are generally unsuitable for establishing semantic equivalence. First, they offer no guarantees of behavioural equivalence, as even minor bytecode differences may alter program behaviour. Second, similarity-based equivalence is inherently non-transitive, limiting its utility for reasoning about correctness-preserving transformations.

Sharma et al.~\cite{sharma2025canonicalization} demonstrate how bytecode canonicalization can improve the success rate of reproducible builds. Building upon our previous work~\cite{dietrich2024levels}, they employ \jnorm for this purpose. Bytecode canonicalization is closely aligned with the concept of equivalence, as both approaches involve comparing transformations, specifically, the \textit{Jimple} files generated by \jnorm, to evaluate the success of a rebuilt artifact. \daleq takes this further by not only establishing equivalence but also offering detailed explanations and formal proofs, thereby increasing the reliability and confidence in the results.


\subsection{Reproducible Builds}

Hassanshahi et al.~\cite{hassanshahi2023macaron} present Macaron, a toolkit aimed at improving software supply chain security. Macaron addresses several challenges related to precise build reproducibility and binary variability by automatically identifying the exact source versions of artifacts and capturing detailed build environment metadata. Although primarily focused on Java/Maven, Macaron also supports additional language ecosystems. Similarly, Keshani et al.\cite{keshani2024aroma} examine methods for identifying the source code corresponding to Maven binary packages. This is a non-trivial task when rebuilding. Those issues led us to include source code analysis into \daleq to check the assumption that the binaries being compared have been built from equal or at least equivalent sources. This will be discussed in Sections~\ref{sec:implementation} and \ref{sec:evaluation:methodology}.

In the broader context, the Linux community has shown sustained interest in reproducible builds, supported by empirical and tooling-focused research. Ren et al.\cite{ren2018automated} and Bajaj et al.~\cite{bajaj2024unreproducible} analyse the prevalence and causes of unreproducible builds in Linux packages. Building on this, Ren et al.~\cite{ren2022automated} introduce RepFix, a tool designed to automatically patch build scripts to improve reproducibility, and demonstrate its effectiveness on a set of Linux packages.

\section{Design}
\label{sec:design}

\subsection{Design Goals}
\label{sec:design:goals}

\Daleq is designed to provide a level 3 equivalence~\cite{dietrich2024levels} for Java bytecode relation. This implies the following:

\begin{enumerate}
	\item The equivalence is a proper equivalence relation, i.e. it is reflexive, symmetric and transitive. 
	\item Equivalent classes should have the same behaviour.
	\item Bytecode sequences representing semantically equivalent instructions are considered equivalent.
	\item Provenance is generated to support both equivalence and non-equivalence statements.
\end{enumerate}

\Daleq establishes equivalence by transforming bytecode and comparing the results of these transformations. Specifically, equivalence is defined as $b_1 \simeq b_2$ if and only if $transf(b_1) = transf(b_2)$. The other three requirements are addressed in Sections \ref{sec:design:soundness}, \ref{sec:design:inference}, and \ref{sec:design:provenance}, respectively.

\subsection{Soundness vs Soundiness}
\label{sec:design:soundness}
 
An equivalence relation should be sound in the sense that it under-approximates (undecidable) behavioural equivalence. 
In other words, equivalent classes should always exhibit the same behavior. However, it is also desirable to establish equivalence for as many class pairs as possible. Note that every pair of artifacts reported as non-equivalent requires manual inspection, which is both costly and prone to error.

In many practical static analyses,  \textit{soundiness} is considered sufficient~\cite{livshits2015defense}. In the context of establishing equivalence between classes, soundy means that classes can be considered equivalent even if their behaviour differs, however, reflection or related features must be used by an application to expose this difference. 
The advantage of accepting soundiness is that a soundy equivalence can create more equivalence statements. From the point of view of an engineer assessing non-equivalent classes for security issues, this equates to removing false positives. This matters are precision is known to be a crucial factor for the acceptance of static analysis tools~\cite{sadowski2018lessons,distefano2019scaling}.

The downside of this approach is that some security issues might be overlooked.  
With \daleq we took the following approach to address this:  \daleq uses a modular design where code normalisation patterns are implemented using separate files consisting of one or several rules, and each rule set is flagged as being either sound or soundy.  Users could therefore simply remove soundy rules if strict soundness was required.  
This also facilitates scenarios where different versions of \daleq with different rule sets could be used simultaneously.

For the soundy rule sets implemented, we carefully assessed the impact those equivalences might have on security.

\subsection{Analysis Pipeline Overview}
\label{sec:design:overview}

\begin{figure*}
        \centering
	\includegraphics[width=\textwidth]{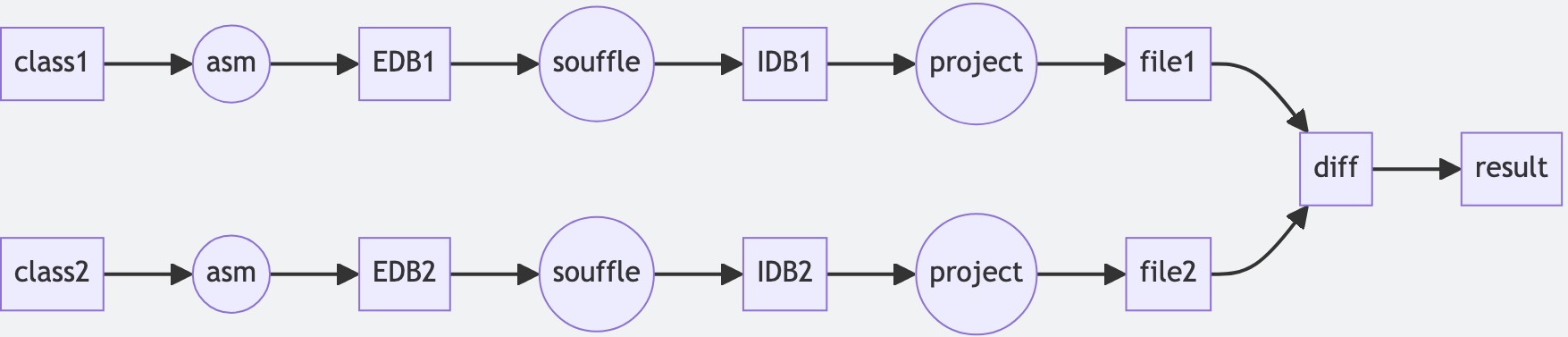}
	\caption{\daleq design overview}
	\label{fig:design-overview}
\end{figure*}

\Daleq disassembles Java byte code and normalises it in three stages: (1) \textbf{extraction}: a low-level relational representation of bytecode is created, this is referred to as the  \textit{EDB} (extensional database) (2) \textbf{inference}:  rules are applied to extract a second database, the \textit{IDB} (intensional database), those rules normalise some bytecode patterns. The rules applied are recorded within the IDB records.  (3) \textbf{projection}: A textual representation of the IDB is created, and auxiliary information is removed from this representation. This representation can then be used for comparison, i.e. to establish equivalence.  Figure~\ref{fig:design-overview} depicts this process.

\subsection{Extraction}
\label{sec:design:extraction}

Extraction is based on \ASM~\cite{bruneton2002asm}. \ASM is a well-established, well-maintained, low-level library used to extract and manipulate Java bytecode.
The extraction layer creates datalog predicates and facts instantiating those predicates.  The predicates fall into two categories: global and instruction predicates. Global predicates represent properties of classes not related to instructions in methods. Examples are superclass, interface, (class) access modifiers, annotations and predicates to list fields and methods declared in a class. 

Instruction  predicates represent bytecode instructions as defined in the JVM spec. For each such instruction, a predicate is defined representing an instruction as a fact. The actual instruction is performed by an instance of a  \texttt{InstructionPredicateFactFactory}  that uses an \ASM node object (from \ASM's tree API) as input. To avoid the performance overhead and limitation on analysability of reflective code, we opted not to use reflection but instead to use a specific factory class for each instruction. As there is a large number of Java bytecode instructions~\footnote{The JavaVM spec version 17 defines 149  \url{https://docs.oracle.com/javase/specs/jvms/se17/html/jvms-6.html#jvms-6.5}}, we generated those factories statically from the respective \ASM AST node types. The respective code generator(s)  are part of the \daleq distribution. 
	
All extracted facts are stored in a tsv text file. There is one such file for each predicate, the name of the file is \texttt{<predicate-name>.facts}. The set of all such facts extracted and a file that defines the respective predicates in \souffle format forms the EDB. This representation of a disassembly is on a level of abstraction similar to the output of the standard Java disassembler \javap.   

Notably, the extraction already applies several normalisations.  Firstly, the EDB resolves constant pool references. For instance, in bytecode, at a call site, the invocation target is represented as a constant pool reference. This is being replaced by a fact that represents the target using a combination of defining class, method name and descriptor. The \ASM API used already provides resolution of constant pool references.  
 
Secondly,  labels are normalised.  \ASM label nodes are mapped to simple names (label1, label2, etc) in EDB records. Label nodes that are never used (as jump targets, e.g. in conditional statements) are ignored.  
 
Thirdly,  line number tables are ignored, and there are no facts in the EDB to associate instructions with line numbers. We consider line numbers as code-reformatting noise during builds and inserting special legal comments into source code files will change those, but has no effect on the semantics of the program.

In terms of equivalence levels~\cite{dietrich2024levels}, this implies that comparing the EDBs of two .class files is a level 2 equivalence. I.e. aspects of bytecode not related to its semantics are removed (line numbers, unused labels), and only isomorphic transformations (label renaming) are applied. 

 Each EDB predicte has a \textit{factid} column of type \textit{symbol} as its first column. Unique values for this column are auto-generated by the extraction, using a simple pattern (``F'' followed by a counter). Those ids are globally unique across predicates.

\subsection{Inference}
\label{sec:design:inference}

The inference layer uses the \souffle datalog engine. Datalog in general, and \souffle in particular, are popular in static program analysis~\cite{whaley2005using,hajiyev2006codequest,smaragdakis2010using,antoniadis2017porting}. This is facilitated by datalog's simple structure and fix point semantics. 

Applying the datalog rules creates a new database, the IDB, (intentional database) from the input EDB.  This is then used as the base for comparison. 
There is a base set of rules that translates each EDB fact into a corresponding IDB fact. 
This involves a new predicate, and two rules. For instance, consider Listing~\ref{listing:idb:basic}.
This defines an IDB predicate \texttt{IDB\_ILOAD}, and a rule that derives facts from the EDB predicate \texttt{ILOAD} (lines 3-4) representing \texttt{ILOAD} instructions found in the bytecode of a class. 
The first terms are the ids of facts. For brevity, those are omitted in Listing~\ref{listing:idb:basic}, we will discuss them later in  Section~\ref{sec:design:provenance}.
The structure of the IDB predicate mirrors the respective EDB predicate.  The  first rule (line 3) translates the EDB fact into an equivalent IDB fact.  This rule uses a second prerequisite (line 5) that acts as a guard. Custom rules can use the \texttt{REMOVED\_INSTRUCTION} predicate to prevent certain IDB facts from being created.   The standard rules for bytecode instructions also instantiate a second generic predicate \texttt{IDB\_INSTRUCTION} (line 6-7). This can be used to describe patterns that apply to all instructions in a certain method. 

\begin{lstlisting}[label=listing:idb:basic,caption=Generated rules and predicate to define the IDB for the ILOAD instruction.]
.decl IDB_ILOAD(factid: symbol,methodid: symbol,instructioncounter: number,var: number)
.output IDB_ILOAD
IDB_ILOAD(..,methodid,icounter,var) :- 
	ILOAD(..,method,icounter,var),
	!REMOVED_INSTRUCTION(_,method,icounter).
IDB_INSTRUCTION(..,method,icounter,"ILOAD") :- 
	ILOAD(..,method,icounter,_).
		
\end{lstlisting}


In addition to the standard rule set that describes EDB and IDB predicates and mappings between those, several custom rules are used. They are all organised in separate .souffle files stored in project resources (i.e. in \texttt{src/main/resources/rules}), and discovered and merged at runtime. 
The first custom rule set is \texttt{access.souffle}.  This is included for convenience and readability, as it maps integer-encoded access flags to readable facts. An example is shown in Listing~\ref{listing:idb:access}, defining facts for \textit{synthetic} elements in a class. The id used references a class, field or method. Note that \textit{band} is an operation provided by \souffle. The premise is an \texttt{IDB\_ACCESS} fact, itself derived from the EBD \texttt{ACCESS} fact with a standard rule.

\begin{lstlisting}[label=listing:idb:access,caption={Custom rule to define synthetic classes, fields or methods}]
.decl IDB_IS_SYNTHETIC(factid: symbol,Id: symbol)
IDB_IS_SYNTHETIC(..,id) :- 
	IDB_ACCESS(factid,id,access), (access band 0x1000)!=0.
.output IDB_IS_SYNTHETIC
\end{lstlisting} 

Other custom rules are related to normalisations, and aim at creating an equivalence IDB even though EDBs extracted from the respective classes are different.
As a first example, consider a rule to normalise the facts representing the bytecode version of a .class file.
This is shown in Listing~\ref{listing:idb:version}. 
Here the bytecode version is always set to 0, enabling bytecodes to be equivalent even if their versions are different.
Note how the guard fact for the predicate \texttt{REMOVED\_VERSION} is instantiated by the second rule (line 4) in order to disable the default rule that would otherwise create a second \texttt{IDB\_VERSION} fact with the version found in bytecode.

\begin{lstlisting}[language=prolog, label=listing:idb:version,caption=Custom rule to ignore the bytecode version.]
.decl IDB_VERSION(fid: symbol,classname: symbol,version: number)
IDB_VERSION(..,classname,0) :- VERSION(fid,classname,_).
.output IDB_VERSION
REMOVED_VERSION(..,classname) :- VERSION(fid,classname,_).
\end{lstlisting} 

In its current version, \daleq contains rules to implement the following normalisation patterns (in addition to the normalisation performed during EDB construction):

\begin{enumerate}
	\item [R1] Null checks in method reference operator, see also~\cite[Pattern N6]{schott2024JNorm}
	\item [R2] Redundant \texttt{checkcast} instructions, see also~\cite[Pattern N9]{schott2024JNorm}
	\item [R3] Inline \texttt{\$values()} method in enumerations, see also~\cite[Pattern N10]{schott2024JNorm}
	\item [R4] Ignore bytecode versions. 
	\item [R5] Invocation type of root methods defined in\texttt{ java.lang.Object} on an object declared using an interface type.~\footnote{This pattern is not included in \cite[Pattern N9]{schott2024JNorm}, see \url{https://github.com/openjdk/jdk/pull/5165} and linked issues for a discussion of the change in the compilation strategy}
	\item [R6] Anonymous inner classes are always implicitly final, but the flag is inconsistently set across different compilers. This is addressed by setting the access modifier of all anonymous inner classes to final~\footnote{\url{https://bugs.openjdk.org/browse/JDK-8161009}}.
\end{enumerate}


The selection of these patterns is based on their frequency of occurrence during the evaluation in the previous work~\cite[Sect. 3]{dietrich2024levels}.
We did not implement all patterns reported by Schott et al and implemented in the \textit{JNorm} tool. We ignored patterns that only apply when  normalising Java byte code produced by \textit{javac} versions prior to Java 8 (N2,N3,N4,N5,N11) as we only rarely encounter them. N12 and N13 are normalisation patterns related to JEP280 and JEP181, respectively.  While those changed apply at Java 11, we did not encounter many instances of those changes causing inequality.  Support for N13 would require some changes to how \daleq works at the moment: reasoning is based on processing one (bytecode) class at a time, and N13 would require to make changes to an entire nest of classes, i.e. the set of classes generated from a single compilation unit, containing a class and all of its inner classes. 

Two other JNorm patterns were rejected due to soundness concerns. Firstly, support for N1 (removal of synthetically generated methods) would introduce unsoundness as synthetic methods are part of the program semantics.  Likewise, support for N7 (buffer method invocation) establishes equivalence for classes that would result in different behaviour for downstream clients. This is more subtle, and relates to binary compatibility. A detailed discussion of this particular issue can be found in~\cite{dietrich2025towards}. 

To support the null checks in N6, we did not just replace calls to \texttt{Object::getClass} by \texttt{Objects.requireNonNull}~\footnote{\textit{JNorm} transforms all occurrences back to the old
null-checking mechanism, we chose the newer pattern to normalise}. The reason is that this could be behaviour-changing as different objects will end up on the stack depending on the method being invoked:  \texttt{Objects.requireNonNull} returns the argument, while  \texttt{Object::getClass} returns an object representing the type of the argument~\footnote{See also \url{https://bugs.openjdk.org/browse/JDK-8073432}}. The respective rules therefore also look for guards to ensure the consistency of the stack -- that the invocation of  \texttt{Object::getClass} is preceded by a \texttt{DUP} and succeeded by a \texttt{POP} instruction.  We find that in the context of method reference operator, the compiler generates such guards. Representing those constraints in datalog is straight-forward.

A particular interesting example is the normalisation of non-null checks (pattern 1), shown in Listing~\ref{listing:idb:nonnull}.  Here an \texttt{invokevirtual Object::getClass} is replaced by an \texttt{invokestatic Objects::requireNonNull}. This is unsound in the sense that it changes the program behaviour as after the respective invocation, different objects are on the stack (the object vs an object representing its class). However, those statements are embedded by the compiler between a \texttt{DUP} and a \texttt{POP} instruction, ensuring the consistency of the stack. The respective rule uses a pattern in the body that reflects this (lines 4 and 5)~\footnote{In this rule, an instruction counter offset is used to identify the previous and the next bytecode instruction. This offset is used during extraction to define instruction counters defining the order of instructions within a method, which is configurable.}.  

\begin{lstlisting}[language=prolog, label=listing:idb:nonnull,caption=Custom rule to normalise non-null checks.]
.decl LEGACY_NON_NULL_CHECK(factid: symbol, method: symbol, icounter: number)
LEGACY_NON_NULL_CHECK(..,method,icounter) :- 
	INVOKEVIRTUAL(factid1,method,icounter,"java/lang/Object","getClass","()Ljava/lang/Class;",_),
	DUP(factid2,method,icounter-100),
	POP(factid3,method,icounter+100).
IDB_INVOKESTATIC(..,method,icounter,"java/util/Objects","requireNonNull","(Ljava/lang/Object;)Ljava/lang/Object;",0) :- 
	LEGACY_NON_NULL_CHECK(factid1,method,icounter).
REMOVED_INSTRUCTION(..,method,icounter) :- 
	LEGACY_NON_NULL_CHECK(factid1,method,icounter).
\end{lstlisting} 

Two rules, R3 and R6 are only soundy as (synthetic) methods (R3)  and final flags (R6) can be queried by programs at runtime via reflection. In both cases, it is highly unlikely that those subtle differences in behaviour can be exploited for security violations.  Note that the R6 only applies to anonymous inner classes. 

In terms of equivalence levels~\cite{dietrich2024levels}, the comparison of the IDBs computed from  two .class files is a level 3 equivalence.  I.e. some abstraction takes place to identify bytecode that does not alter the behaviour of the program. 

\begin{figure}
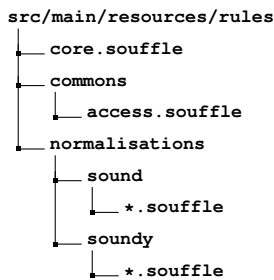

	\begin{scriptsize}
	\dirtree{%
		.1 \textbf{src/main/resources/rules}.
		.2 \textbf{core.souffle}.    
		.2 \textbf{commons}.   
		.3 \textbf{access.souffle}.
		.2 \textbf{normalisations}.
		.3 \textbf{sound}.
		.4 \textbf{*.souffle}.  
		.3 \textbf{soundy} .
		.4 \textbf{*.souffle}.
	}
	\end{scriptsize}
	\caption{Rule folder structure}.
	\label{fig:rule-structure}
\end{figure}

Rules are organised in a folder structure depicted in Figure~\ref{fig:rule-structure}. 
They are loaded by the class loader, allowing third parties to provide additional rules.

\subsection{Recording Datalog Inference}
\label{sec:design:provenance}

The inference rules record provenance through the aggregated ids in the first term (column) of each fact. For facts,  unique identifiers (``F1'', ``F2'', ..) are generated during extraction, and used in the first term in each fact. When a rule is applied and a new fact is inferred, a composite identifier is created.
For instance, in Listing~\ref{listing:idb:version}, the id terms in rule heads were omitted for brevity. For the first rule (line 2), this term is \texttt{cat("R\_REMOVE\_BYTECODE\_VERSION","[",fid,"]")}. If the id of the fact \textit{fid} in the premise was \texttt{F42}, then this would create an id \texttt{R\_REMOVE\_BYTECODE\_VERSION[F42]} for the derived fact.


The constructed ids of derived facts encode the derivations that were used to construct them. I.e., those values can be parsed and presented as a derivation tree. \daleq includes a simple grammar for the embedded provenance language, along with utilities to parse these expressions and render them as trees in the final HTML report, which serves as the tool's output. This grammar is shown in Listing~\ref{listing:idb:grammar}.

\begin{lstlisting}[label=listing:idb:grammar,caption=Grammar for constructed ids encoding derivations.]
grammar Proof;
proof : node <EOF> ;
node : ID children? ;
children : '[' node (',' node)* ']' ;
ID : [a-zA-Z0-9_]+ ;
\end{lstlisting} 

Using those derivations, we can provide provenance to users supporting equivalence statements. 
For non-equivalence statements, standard diff tools can be employed. We will discuss this in the following section.

\subsection{Projection and Establishing Equivalence}
\label{sec:design:projection}

This is the last step of the analysis pipeline.
While the IDBs are the base of comparing bytecodes, they cannot be used directly for two reasons. Firstly, the ids of derived facts contain information about the derivation, i.e. \textit{how} the normalisation was achieved. To compare normalised code, this information needs to be removed.  Secondly, the IDB still contains instruction counters. Those counters are only used to define the order of instructions. I.e. the particular numbering scheme is irrelevant as long as the order is retained. 
We therefore remove the respective terms from facts representing bytecode instructions.

The result of those two steps is referred as the \textit{projection}. To establish equivalence, the IDB projections are printed into a textual representation, and then compared. If they are the same, equivalence has been established. Using a textual representation facilitates the use of standard diff tools. If classes are not equivalent, the produced diffs can be used as provenance. If they are equivalent, provenance is provided by a report that includes the derivations of rules applied.

\section{Implementation}
\label{sec:implementation}

\subsection{CLI and Report Generation}

\Daleq consists of several components. The extractor, which generates the EDB, is implemented in Java, while the normalisation rules are written in the \souffle datalog~\footnote{https://souffle-lang.github.io/docs.html}. \daleq is primarily designed as a command-line interface (CLI) tool, but it can also be integrated into CI/CD pipelines for artifact validation. The tool accepts two jar files as input and generates an HTML report. Additionally, it can accept jar files containing source code, alongside the bytecode jars, to verify that the compared bytecodes were built from the same source code. This leverages the fact that the source code used for building the binaries is often distributed with them.

\begin{figure*}
        \centering
	\includegraphics[width=\textwidth]{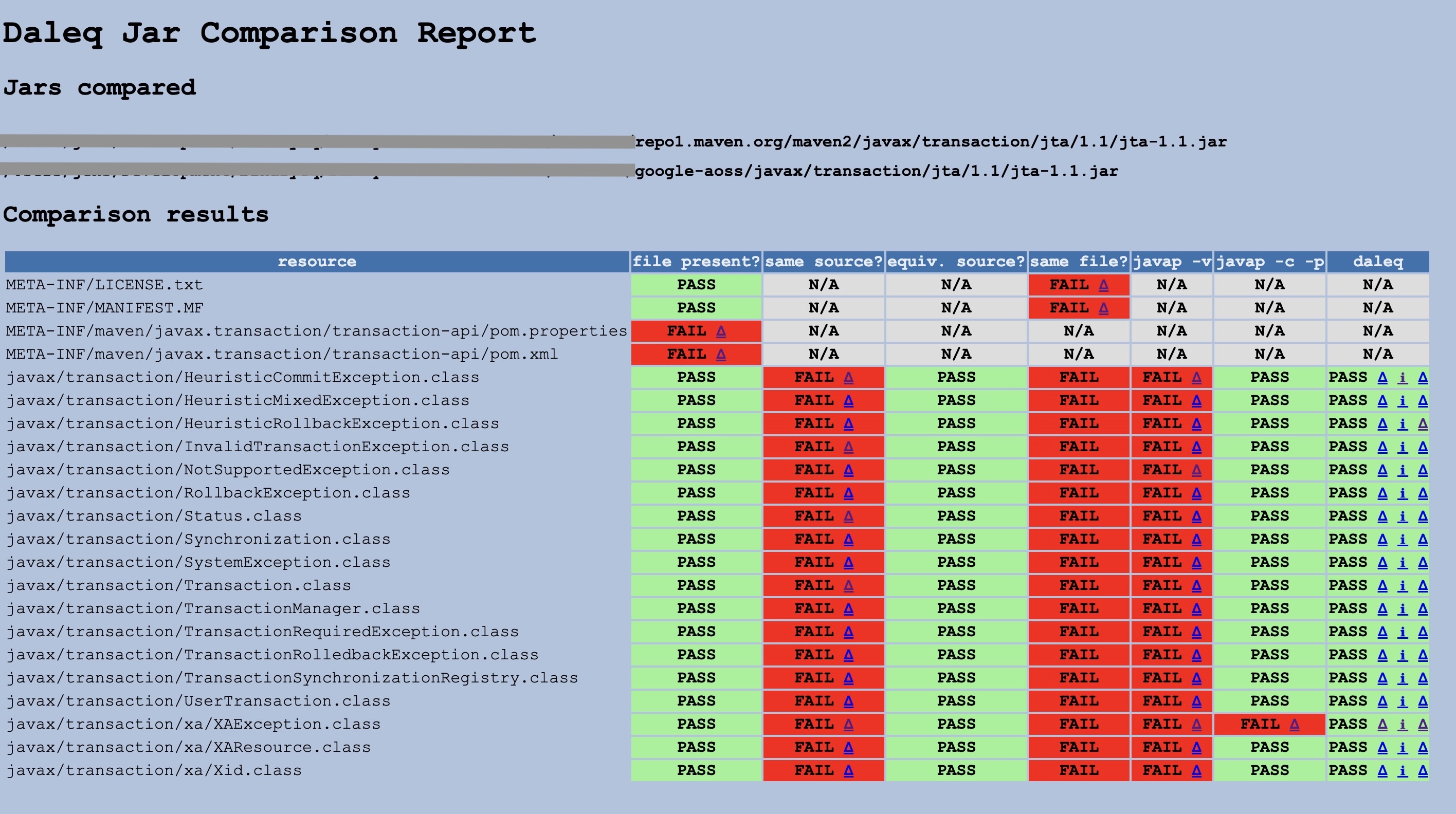}
	\caption{Report generated by \daleq}
	\label{fig:ui}
\end{figure*}

An example of an HTML report output is shown in Figure~\ref{fig:ui}. Here, we compare the jar file for the artifact \textit{javax.transaction:jta:1.1} from Maven Central with the corresponding artifact rebuilt by \gaoss.

\Daleq employs various \textit{analysers} to compare classes, metadata and resources within the jars. The results are displayed in a table, the rows correspond to files within the jar(s), and the columns correspond to the various analysers. Possible analysis result states are:

\begin{itemize}
	\item \textbf{PASS} the comparison yields true 
	\item \textbf{FAIL} the comparison yields false
	\item \textbf{N/A }the comparison cannot be applied (e.g., a source code comparison cannot be used to compare bytecode)
	\item \textbf{ERROR} the comparison has resulted in an error
\end{itemize}

The result states in the report are similar to the ones used by automated testing frameworks like \textit{JUnit}~\footnote{In testing, often \textit{skip} is used instead of \textit{N/A}}. Small markers next to the results indicate that they are explainable; these markers link to additional pages with more details. For failed comparisons, the linked pages typically display diffs (rendered in HTML), while for errors, they contain error logs.

The report lists all files found in either jar. The first check (column 2) verifies that the file is present in both jars. It then compares the sources for equality and equivalence~\footnote{Using an AST-based analysis that ignores formatting and comments} (columns 3 and 4). In the example shown in Figure~\ref{fig:ui}, there are two resources in \texttt{META-INF/maven} that are present in the \gaoss but not in the \mvnc built jar. Interestingly, the sources used to create the jars are not identical. A closer inspection of the respective diffs reveals that this is caused by altered legal comments. But sources are still shown to be equivalent by the respective analysis. The $\Delta$ link references a page showing the diff generated during the analysis. 

The same file comparison (column 5) then compares the content of the respective files, this reveals some changes in the metadata files \texttt{META-INF/LICENSE.txt} and \texttt{META-INF/MANIFEST.MF}. 

The last two columns present the results of two bytecode analyses: one based on the standard disassembler (i.e., comparing the output of \texttt{javap -c -p}) and the other from the \daleq comparison. \Daleq generates several reports to facilitate explainability, including links to the EDBs and IDBs used in the analysis.

Notably, \javap shows equivalence for all classes except \texttt{XAException.class}, where the disassemblies differ. Details are provided in the linked diff created from the disassemblies. However, \daleq can still establish equivalence between the two versions of this class.

\subsection{Explainability}

For \daleq equivalence results, \textit{advanced-diff} reports are generated and made accessible through links in the main report. These reports include all extracted and derived facts, along with explanations of how custom rules were applied to establish equivalence. An example of this section of the report is shown in Figure~\ref{fig:ui}. Derivation trees are visualized using CSS, and users can click on the facts and rules involved in the derivation.

\subsection{Adding Additional Analyses}

The \daleq tool features an extensible design, allowing new analyses to be added via the \texttt{io.github.bineq.daleq.cli.Analyser} interface, which can be implemented by third parties. Analysers are discovered and loaded through the Java service loader mechanism, enabling them to be used as plugins.

\begin{figure}
	\includegraphics[width=\columnwidth]{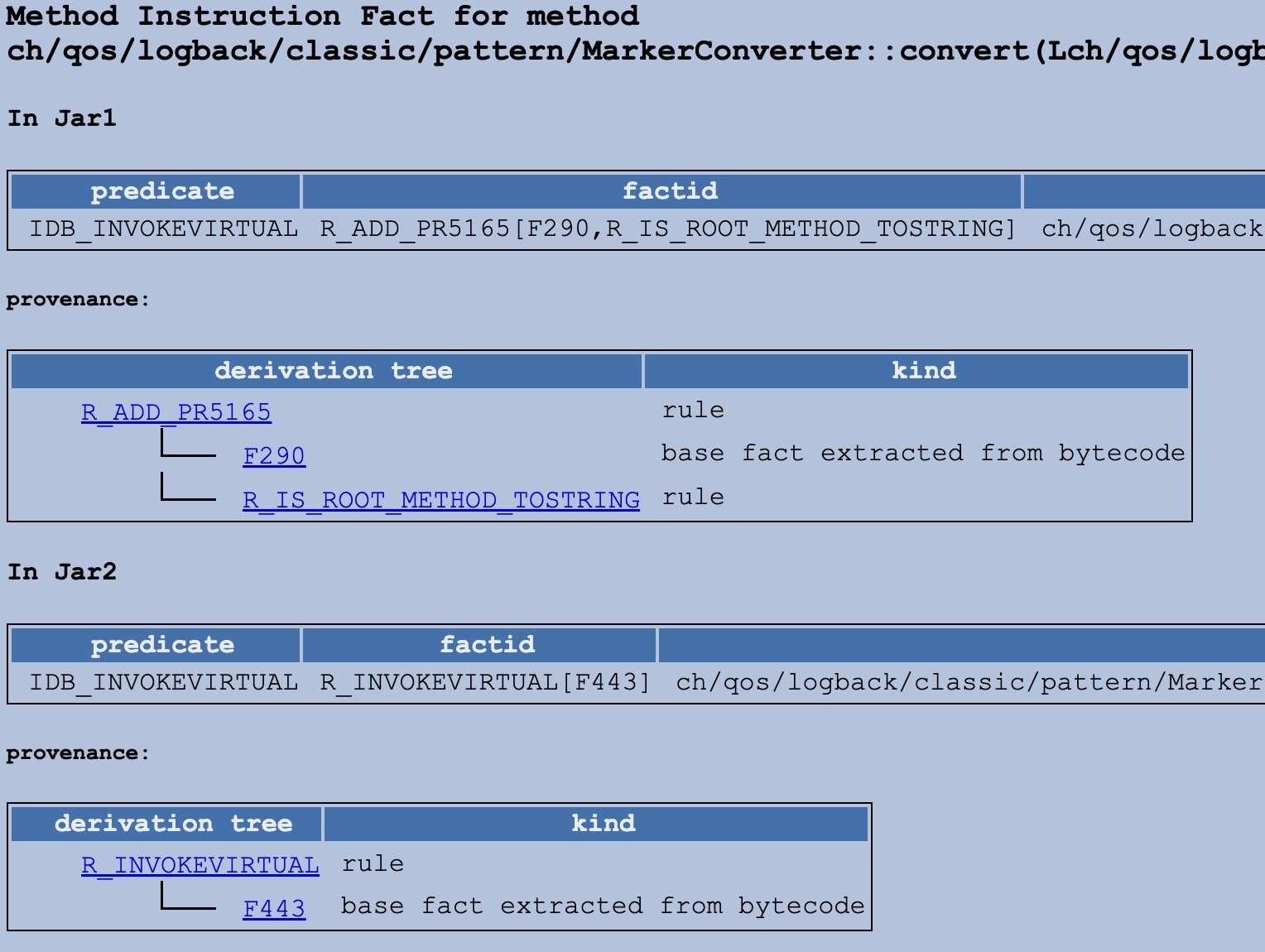}
	\caption{Provenance user interface}
	\label{fig:provencance-ui}
\end{figure}
 
\section{Evaluation}
\label{sec:evaluation}
\label{sec:evaluation:results}



The evaluation is guided by the following research questions:

\begin{itemize}
    \item [\textbf{RQ1}] How common are bitwise differences in Java class files between developer-built artifacts (\mvnc) and independently rebuilt artifacts from trusted providers such as \bfs and \gaoss?

     \item [\textbf{RQ2}] How effective is \daleq in classifying non-bitwise-equal classes as equivalent compared to existing tools such as \javap and \jnorm?

     \item [\textbf{RQ3}] What are the most frequent causes of non-equivalence in rebuilt Java class files, and to what extent can they be explained through recurring bytecode-level patterns?
     
      \item [\textbf{RQ4}]  How scalable is \daleq?


\end{itemize}  

\subsection{Methodology}
\label{sec:evaluation:methodology}

To evaluate the performance of \daleq, we used the dataset from~\cite{dietrich2024levels}~\footnote{Available at \url{https://zenodo.org/records/14915249}}.
In particular, we compared jars built by the developer (deployed on Maven Central \mvnc) with alternatively built jars from Oracle's Build-From-Source (\bfs) and Google's Assured Open Source (\gaoss) projects. 
We ensured that the respective sources (released with the jars) are equivalent, using the AST-based comparison also used in ~\cite{dietrich2024levels}.  The number of jars and classes compared are summarised in Table~\ref{tab:non-equal-classes}. Both datasets \mvncvsbfs and \mvncvsgaoss contain a similar number of classes (i.e. compiled class files, *.class), ca 130,000.  The numbers of jars significantly differs, \bfs contains more smaller jars, whereas \gaoss contains fewer but larger jars. In both cases, we found a small number of jars without .class files. These jars contain only resources and meta-data, and were therefore excluded from the analysis.

We have created scripts that summarise the results for RQ1 and RQ2. They also produce a file structure,
which is used as input for additional analysis scripts that answer RQ3 and RQ4. The corresponding data is available at \url{https://zenodo.org/records/16628896}.

\subsection{RQ1 Results}

Results for RQ1 can be found in Table~\ref{tab:non-equal-classes}. 
We found a significant number of non-equal classes across alternative builds, 2,158 (1.58\%) for the \mvncvsbfs comparison, and 33,470 (25.69\%) for the \mvncvsgaoss comparison.  

Interestingly, different processes and design goals by alternative providers lead to significantly different rates of classes that are bitwise identical. 

\begin{table*}[!ht]
	\centering
	\begin{tabular}{|l|l|l|l|l|l|l|l|}
		\hline
		provider1 & provider2 & jars & jars without classes & classes & equal classes & non-equal classes & non-equal classes (\%) \\ \hline
		\mvnc & \bfs & 1,922 & 22 & 135,425 & 133,267 & 2,158 & 1.59\% \\ 
		\mvnc & \gaoss & 792 & 13 & 130,265 & 96,795 & 33,470 & 25.69\% \\ \hline
	\end{tabular}
	\caption{Non-equal classes}
	\label{tab:non-equal-classes}
\end{table*}

\subsection{RQ2 Results}

We then identified classes that are not bitwise equivalent (i.e., level 1), but can still be shown to be equivalent at higher levels. To evaluate the performance of \daleq relative to other tools, we used the standard Java disassembler~\footnote{\texttt{javap -c -p}} and \jnorm~\footnote{version 1.0.0 with the normalisation option \texttt{-n}.}. \jnorm  was the best-performing tool in a previous study~\cite{dietrich2024levels}. 
Note that in \cite{dietrich2024levels} \jnorm was also used with the aggressive normalisation option. We did not use aggressive normalisation here as it includes some normalisation patterns that we consider unsound, and even unsoundy, such as N16 described in~\cite{schott2024JNorm}. 
\javap was also used in \cite{dietrich2024levels} and performed reasonably well. In particular, it was able to analyse all classes, whereas some of the other tools like the \textit{fernflower} decompiler (used in  \cite{dietrich2024levels}) and \jnorm (used both in  \cite{dietrich2024levels} and this study) sometimes resulted in errors. \javap is also a tool that is trusted by engineers as it is well-maintained, and part of the standard Java Developer Kit. It therefore forms a suitable baseline of the evaluation of tools to establish equivalence.  Using the \texttt{-c -p} configuration provides some abstractions, whereas using \texttt{-v} (verbose) would have resulted in a very detailed representation (including the line number table and the constant pool) not suitable to establish equivalence between different classes. 

Equivalences are established by comparing bytecode transformations generated by the respective tool. Those are then diffed, and if the diff is empty, the respective classes are considered equivalent. Otherwise the diff is stored, those diffs are then used later to answer RQ3.  

The results are shown in Table~\ref{tab:equivalent-classes}. \Daleq can infer that 85.91\% of non-equal classes in the \mvncvsbfs comparison and 90.80\% of non-equal classes in the \mvncvsgaoss comparison can be shown to be equivalent.  This indicates a substantial reduction in manual effort required to assess these cases. \Daleq   outperforms both the \javap - (45.18\% / 35.08\%) and the \jnorm - based (65.06\% / 80.90\%) equivalences. There were also cases where \jnorm encountered errors, which are reported in the last column and counted as \jnorm-non-equivalent. In contrast, \daleq successfully analyzed all classes in the evaluation dataset.

\begin{table*}[!ht]
	\centering
	\begin{tabular}{|l|l|l|l|l|l|l|}
		\hline
		provider1 & provider2 & non-equal classes & equiv. (javap) & equiv. (jnorm) & equiv. (daleq) & errors (jnorm) \\ \hline
		\mvnc & \bfs & 2,158 & 975 (45.18\%) & 1,404 (65.06\%) & 1,854 (85.91\%) & 5  (0.23\%)\\ 
		\mvnc & \gaoss & 33,470 & 11,741  (35.08\%) & 27,079 (80.90\%) & 30,390 (90.80\%) & 461 (1.38\%) \\ \hline
	\end{tabular}
	\caption{Equivalent classes}
	\label{tab:equivalent-classes}
\end{table*}

\subsection{RQ3 Results}
\label{sec:evaluation:non-equivalent}


While \daleq generally performs well and significantly reduces the manual effort required to assess non-equivalent classes, a substantial number of such classes remain in the report. We expect that additional patterns can be discovered and implemented using \souffle normalisation rules, and we make no claim that the current set of rules is exhaustive.


To gain a better understanding of the nature of non-equivalent classes, we analysed them using the following approach.
If two .class files are not equivalent, a diff file \texttt{daleq-diff.txt} is generated showing the differences of the respective IDBs in standard diff format. 
There are 3,384 such files~\footnote{This is the sum of non-\daleq equivalent records reported in Table~\ref{tab:equivalent-classes}, i.e. (2,158-1,854)+(33,470-30,390)} in the results. 
We analysed these files to identify common patterns in changes.
Pattern detection was implemented using a text analysis of added and removed lines. For instance, to check for changes in loaded constants, we collected lines starting with \texttt{+IDB\_LDC} and \texttt{-IDB\_LDC}, respectively. Then we extracted the values loaded by collecting the last tokens from those lines (the lines represent database records, using a tab as separator), compared those to sets, and report an instance of this pattern if those sets are different. This can potentially produce some false negatives, e.g. if constants loaded are swapped between different methods. We consider this unlikely.  



The following are the analysed patterns:

\noindent \textbf{CHECKCAST} A checkcast statement is removed or added.  In general, checkcasts can change the behaviour of programs  -- when checkcasts fail, a runtime exception is created, changing the control flow of a program. Casts can therefore not be ignored, unless behavioural equivalence can be inferred from context~\footnote{As is the case in P2 discussed in Section~\ref{sec:design:inference}}. An example is shown in Listing~\ref{listing:diff}. This issue is also discussed in~\cite{schott2024JNorm} (N16). The authors argue that this is not used in normalisation, but only in aggressive normalisation mode.  

\begin{lstlisting}[label=listing:diff,caption=Non-equivalence caused by an additional \texttt{checkcast} instruction (some lines abbreviated). ]
--- version1
+++ version2
@@ -582,6 +582,7 @@
 IDB_GETFIELD	..
 IDB_ALOAD	..
 IDB_INVOKEVIRTUAL	..
+IDB_CHECKCAST	projected	org/apache/curator/shaded/com/google/common/graph/StandardNetwork::incidentNodes(Ljava/lang/Object;)Lorg/apache/curator/shaded/com/google/common/graph/EndpointPair;	-1	org/apache/curator/shaded/com/google/common/graph/NetworkConnections
 IDB_INVOKESTATIC	..
 IDB_CHECKCAST	..
 IDB_ALOAD	..
		
\end{lstlisting}

\noindent \textbf{CONSTANT} A  constant value in a LDC instruction is changed.  We found instances where different values point to different paths. This can have security implications, for instance, differences pointing to file system locations may indicate path traversal (CWE-22) attacks. We found several path references that differ between \mvnc and \gaoss builds of \textit{dev.zio:zio-test\_3:2.1.0-RC2}, those are references to scala source files, the paths seem to point to the build environment being used, starting with \textit{/home/runner/work/} and \textit{/workspace/shared-workspace/build/upstream-repo}, respectively. 


\noindent \textbf{SBINIT} \texttt{StringBuilder} initialisation.  There are different sequences being generated to initialise  \texttt{java.lang.StringBuilder}, using different descriptors \texttt{()V} and \texttt{(I)V}, respectively.  

\noindent \textbf{SIGNTR} Missing signature attributes in methods. There are cases where one class misses the signature. This is a candidate for another "soundy" rule, those changes may effect program behaviour if reflection is used.  

\noindent \textbf{SYNMET} Naming of synthetic methods. Different compilers sometimes name synthetic classes differently.  Ignoring them would make the analysis unsound as those methods encode program behaviour.

\noindent \textbf{SYNFLD} Naming of synthetic fields. Different compiler sometimes name synthetic classes differently.  Ignoring them would make the analysis unsound as those methods fields have an impact on program behaviour. 

\noindent \textbf{ANNO} Changed annotations. In particular, we found subtle differences in the parameters of checker framework's \texttt{@Nullable} annotations.  We consider annotations critical for security as modern frameworks heavily rely on them to define the semantics of a program. For instance, both \textit{springframework} and \textit{micronaut} use annotations to define entry points for web applications. 

\noindent \textbf{ACCESS} Changed access. While we capture a particular pattern (P6), there are other cases where the flags associated with class, methods and fields generated by different builds differ. 

The results of this analysis are summarised in Table~\ref{tab:diffanalysis}, column 2. We also counted the number of classes where non-equivalence has more than one cause in the last row.

\begin{table}[]
	\begin{tabular}{|llll|}
		\hline
		cause     & all           & javap non-equiv.     & jnorm non-equiv.     \\ \hline
		total     & 3,384         & 49           & 523           \\ 
		CHECKCAST & 635 (18.76\%) & 0 (0.00\%)   & 17 (3.25\%)   \\
		CONSTANT  & 152 (4.49\%) & 0 (0.00\%)   & 0 (0.00\%)    \\
		SBINIT    & 650 (19.21\%) & 0 (0.00\%)   & 6 (1.15\%)    \\
		SIGNTR    & 396 (11.70\%)& 0 (0.00\%)   & 385 (73.61\%) \\
		SYNMET & 309 (9.13\%) & 0 (0.00\%)   & 61 (11.66\%)) \\
		SYNFLD   & 612 (18.09\%)   & 0 (0.00\%)   & 0 (0.00\%)    \\
		ANNO      & 75 (2.22\%)  & 45 (91.84\%) & 60 (11.47\%)   \\
		ACCESS    & 630 (18.62\%) & 4 (8.16\%)   & 19 (3.63\%)  \\
		multiple  & 926 (27.36\%) & 0 (0.00\%)   & 20 (3.82\%) \\ \hline
	\end{tabular}
	\caption{Analysis of pairs of classes not \daleq-equivalent}
	\label{tab:diffanalysis}
\end{table}

At least two of those patterns (SIGNTR and ACCESS) are candidates for additional soundy rules.  It is not clear whether normalising synthetic method names is expressible in \daleq, if so, some of the differences in the SYNFLD and SYNMET categories could also be covered by additional rules. CONSTANT is category that could indicate a compromised binary, and warrants further investigation.  

We then analysed \daleq non-equivalent classes flagged as equivalent by either \javap or \jnorm. 
The results are also included in Table~\ref{tab:diffanalysis}, columns 3 and 4, respectively. 
We note here that the \javap  equivalence is based on running \texttt{javap -c -p} which produces a high-level representation of bytecode with some abstractions. Had we used  \texttt{javap -v}, those values would have been much lower, possibly zero. On the other hand, the results for \javap as reported in Table~\ref{tab:equivalent-classes} would have been worse, too. Those choices reflect the classic tradeoff between precision and recall.  There are actual cases where the javap-based equivalence becomes unsound, such an example (changed value of constant not used in the defining class) was detected in discussed in~\cite{dietrich2024levels}.

For \jnorm, we note here that \jnorm has not been designed to establish byte code equivalence, but for \textit{code similarity analysis}. For instance, the issues caused by synthetic methods (SYNMET) are a result of \jnorm ignoring those methods (normalisation rule N1 in~\cite{schott2024JNorm}), which we consider unsound.  We also found that the jimple representation used by jnorm sometimes ommits information like signature headers, leading to the high numbers in the SIGNTR category. 


\subsection{RQ4 Results}

To assess the scalability of \daleq, we have computed some statistics from the data captured in the \textit{computation-time-in-ms.txt} files
Those files record the time taken to extract the EDB, compute the IDB and project it. The experiments where performed on an Apple M1 Pro running MacOS 15.5 with 32GB memory. The JVM used was the \textit{OpenJDK Runtime Environment Corretto-21.0.5.11.1}, the \souffle version used was 2.4.1.

We analysed 71,256 such  timestamps. This number is twice the sum the number of non-equivalent classes (see Table~\ref{tab:non-equal-classes}, column 7). For equal (i.e. bitwise identical) classes, \daleq-equivalence is not computed but inferred. This hinges on the assumption that \daleq and the tools it relies on (\ASM , \daleq) are deterministic, i.e. the same EDBs and IDBs are computed from identical input.

The mean time it takes to compute the projected IDB for a class is 1,691 ms.  
To put this in context, consider a larger Java library like \textit{guava-33.4.0-jre.jar}, containing 2,018 .class files.
Further assume that of those classes 20\% are not bitwise identical. This means that for 808 classes (2 * 0.2 *  2,018) \daleq has to compute the projected IDB.  This would take under 23 mins. Some additional time would be required to compare the generated reports for equality, and diffs resources and metadata, but those operations are very fast. 
Most libraries are significantly smaller than this, and a typical \daleq analysis takes only a few minutes.

\section{Conclusion}
\label{sec:conclusion}

In this paper, we have introduced \daleq, a tool designed to compare Java bytecode and establish equivalence, facilitating the assessment of rebuilt artifacts. \daleq operates at level 3 equivalence, with a particular focus on providing provenance for both equivalence and non-equivalence statements. This is achieved through a datalog-based implementation of bytecode normalisations that records derivations and exposes them to users, enhancing transparency and traceability. The evaluation on a large real-world dataset further confirms that \daleq performs well and outperforms the state-of-the-art \jnorm tool. 

We have demonstrated the impact of \daleq in an industrial context by evaluating artifacts built from source by two vendors. The results show a significant reduction in the manual effort required to assess non-bitwise equivalent artifacts, which would otherwise require intensive human inspection. \daleq’s ability to automatically establish equivalence and explain the reasoning behind these results offers substantial time savings, especially when dealing with large-scale datasets.

\Daleq is publicly available. By releasing it to the public, we invite contributions to enhance its functionality. We anticipate that this will lead to the development of additional rules and integration of other analysers into the tool.


\subsection{Tools and Data Availability}

\begin{table}[h!]
	\begin{tabular}{rl}
		tool:                   & \url{https://github.com/binaryeq/daleq/}           \\
		evaluation scripts:     & \url{https://github.com/binaryeq/daleq-evaluation} \\
		input dataset:          & \url{https://zenodo.org/records/14915249}          \\
		evaluation result data: & \url{https://zenodo.org/records/16628896}        
	\end{tabular}
\end{table}

\section*{Acknowledgments}

The work of the first author was supported by a gift by Oracle Labs Australia.

\bibliographystyle{plain}
\bibliography{references}

\begin{thebibliography}{10}

\bibitem{reproduciblebuild}
Reproducible builds.
\newblock \url{https://reproducible-builds.org/}.

\bibitem{xz}
{CVE-2024-3094 (xz)}, 2024.
\newblock \url{https://nvd.nist.gov/vuln/detail/CVE-2024-3094}.

\bibitem{andreoli2023prevalence}
Anthony Andreoli, Anis Lounis, Mourad Debbabi, and Aiman Hanna.
\newblock On the prevalence of software supply chain attacks: Empirical study
  and investigative framework.
\newblock {\em Forensic Science International: Digital Investigation},
  44:301508, 2023.

\bibitem{antoniadis2017porting}
Tony Antoniadis, Konstantinos Triantafyllou, and Yannis Smaragdakis.
\newblock Porting doop to souffl{\'e}: a tale of inter-engine portability for
  datalog-based analyses.
\newblock In {\em Proceedings of the 6th ACM SIGPLAN International Workshop on
  State Of the Art in Program Analysis}, pages 25--30, 2017.

\bibitem{bajaj2024unreproducible}
Rahul Bajaj, Eduardo Fernandes, Bram Adams, and Ahmed~E Hassan.
\newblock Unreproducible builds: Time to fix, causes, and correlation with
  external ecosystem factors.
\newblock {\em Empirical Software Engineering}, 29(1):11, 2024.

\bibitem{baker1998deducing}
Brenda~S Baker and Udi Manber.
\newblock Deducing similarities in java sources from bytecodes.
\newblock In {\em USENIX Annual Technical Conference}, pages 179--190, 1998.

\bibitem{bruneton2002asm}
Eric Bruneton, Romain Lenglet, and Thierry Coupaye.
\newblock Asm: a code manipulation tool to implement adaptable systems.
\newblock {\em Adaptable and extensible component systems}, 30(19), 2002.

\bibitem{butler2023business}
Simon Butler, Jonas Gamalielsson, Bj{\"o}rn Lundell, Christoffer Brax, Anders
  Mattsson, Tomas Gustavsson, Jonas Feist, Bengt Kvarnstr{\"o}m, and Erik
  L{\"o}nroth.
\newblock On business adoption and use of reproducible builds for open and
  closed source software.
\newblock {\em Software Quality Journal}, 31(3):687--719, 2023.

\bibitem{chen2014achieving}
Kai Chen, Peng Liu, and Yingjun Zhang.
\newblock Achieving accuracy and scalability simultaneously in detecting
  application clones on android markets.
\newblock In {\em Proceedings of the 36th International Conference on Software
  Engineering}, pages 175--186, 2014.

\bibitem{davis2010whence}
Ian~J Davis and Michael~W Godfrey.
\newblock From whence it came: Detecting source code clones by analyzing
  assembler.
\newblock In {\em 2010 17th Working Conference on Reverse Engineering}, pages
  242--246. IEEE, 2010.

\bibitem{dietrich2014broken}
Jens Dietrich, Kamil Jezek, and Premek Brada.
\newblock Broken promises: An empirical study into evolution problems in java
  programs caused by library upgrades.
\newblock In {\em 2014 Software Evolution Week-IEEE Conference on Software
  Maintenance, Reengineering, and Reverse Engineering (CSMR-WCRE)}, pages
  64--73. IEEE, 2014.

\bibitem{dietrich2024levels}
Jens Dietrich, Tim White, Behnaz Hassanshahi, and Paddy Krishnan.
\newblock Levels of binary equivalence for the comparison of binaries from
  alternative builds.
\newblock In {\em Accepted for ICSME'25}. IEEE, 2025.

\bibitem{dietrich2025towards}
Jens Dietrich, Tim White, Valerio Terragni, and Behnaz Hassanshahi.
\newblock Towards cross-build differential testing.
\newblock In {\em Proc. ICST'25}. IEEE, 2025.

\bibitem{distefano2019scaling}
Dino Distefano, Manuel F{\"a}hndrich, Francesco Logozzo, and Peter~W O'Hearn.
\newblock Scaling static analyses at facebook.
\newblock {\em Communications of the ACM}, 62(8):62--70, 2019.

\bibitem{drexel2024reproducible}
Joshua Drexel, Esther H{\"a}nggi, and Iy{\'a}n~M{\'e}ndez Veiga.
\newblock Reproducible builds and insights from an independent verifier for
  arch linux.
\newblock In {\em Sicherheit 2024}, pages 243--257. Gesellschaft f{\"u}r
  Informatik eV, 2024.

\bibitem{ellison2010evaluating}
Robert~J Ellison, John~B Goodenough, Charles~B Weinstock, and Carol Woody.
\newblock Evaluating and mitigating software supply chain security risks.
\newblock {\em Software Engineering Institute, Tech. Rep. CMU/SEI-2010-TN-016},
  2010.

\bibitem{enck2022top}
William Enck and Laurie Williams.
\newblock Top five challenges in software supply chain security: Observations
  from 30 industry and government organizations.
\newblock {\em IEEE Security \& Privacy}, 20(2):96--100, 2022.

\bibitem{fourne2023s}
Marcel Fourn{\'e}, Dominik Wermke, William Enck, Sascha Fahl, and Yasemin Acar.
\newblock It’s like flossing your teeth: On the importance and challenges of
  reproducible builds for software supply chain security.
\newblock In {\em 2023 IEEE Symposium on Security and Privacy (SP)}, pages
  1527--1544. IEEE, 2023.

\bibitem{hajiyev2006codequest}
Elnar Hajiyev, Mathieu Verbaere, and Oege De~Moor.
\newblock Codequest: Scalable source code queries with datalog.
\newblock In {\em European Conference on Object-Oriented Programming}, pages
  2--27. Springer, 2006.

\bibitem{hassanshahi2023macaron}
Behnaz Hassanshahi, Trong~Nhan Mai, Alistair Michael, Benjamin Selwyn-Smith,
  Sophie Bates, and Padmanabhan Krishnan.
\newblock Macaron: A logic-based framework for software supply chain security
  assurance.
\newblock In {\em Proc. SCORED'23}, 2023.

\bibitem{jayasuriya2023breaking}
Dhanushka Jayasuriya, Valerio Terragni, Jens Dietrich, Samuel Ou, and Kelly
  Blincoe.
\newblock Understanding breaking changes in the wild.
\newblock In {\em ACM SIGSOFT International Symposium on Software Testing and
  Analysis (ISSTA)}, pages 1433–--1444, 2023.

\bibitem{ji2008plagiarism}
Jeong-Hoon Ji, Gyun Woo, and Hwan-Gue Cho.
\newblock A plagiarism detection technique for java program using bytecode
  analysis.
\newblock In {\em 2008 third international conference on convergence and hybrid
  information technology}, volume~1, pages 1092--1098. IEEE, 2008.

\bibitem{keivanloo2014sebyte}
Iman Keivanloo, Chanchal~K Roy, and Juergen Rilling.
\newblock Sebyte: Scalable clone and similarity search for bytecode.
\newblock {\em Science of Computer Programming}, 95:426--444, 2014.

\bibitem{keshani2024aroma}
Mehdi Keshani, Tudor-Gabriel Velican, Gideon Bot, and Sebastian Proksch.
\newblock Aroma: Automatic reproduction of maven artifacts.
\newblock {\em Proc. FSE'24}, (FSE), 2024.

\bibitem{lamb2021reproducible}
Chris Lamb and Stefano Zacchiroli.
\newblock Reproducible builds: Increasing the integrity of software supply
  chains.
\newblock {\em IEEE Software}, 39(2):62--70, 2021.

\bibitem{livshits2015defense}
Benjamin Livshits, Manu Sridharan, Yannis Smaragdakis, Ond{\v{r}}ej Lhot{\'a}k,
  J~Nelson Amaral, Bor-Yuh~Evan Chang, Samuel~Z Guyer, Uday~P Khedker, Anders
  M{\o}ller, and Dimitrios Vardoulakis.
\newblock In defense of soundiness: A manifesto.
\newblock {\em Communications of the ACM}, 58(2):44--46, 2015.

\bibitem{macho2021nature}
Christian Macho, Stefanie Beyer, Shane McIntosh, and Martin Pinzger.
\newblock The nature of build changes: An empirical study of maven-based build
  systems.
\newblock {\em Empirical Software Engineering}, 26(3):32, 2021.

\bibitem{martinez2021software}
Jeferson Mart{\'\i}nez and Javier~M Dur{\'a}n.
\newblock Software supply chain attacks, a threat to global cybersecurity:
  Solarwinds’ case study.
\newblock {\em International Journal of Safety and Security Engineering},
  11(5):537--545, 2021.

\bibitem{mukherjee2021fixing}
Suchita Mukherjee, Abigail Almanza, and Cindy Rubio-Gonz{\'a}lez.
\newblock Fixing dependency errors for python build reproducibility.
\newblock In {\em Proceedings of the 30th ACM SIGSOFT international symposium
  on software testing and analysis}, pages 439--451, 2021.

\bibitem{ohm2020backstabber}
Marc Ohm, Henrik Plate, Arnold Sykosch, and Michael Meier.
\newblock Backstabber’s knife collection: A review of open source software
  supply chain attacks.
\newblock In {\em Proc DIMVA'20}. Springer, 2020.

\bibitem{raemaekers2014semantic}
Steven Raemaekers, Arie Van~Deursen, and Joost Visser.
\newblock Semantic versioning versus breaking changes: A study of the maven
  repository.
\newblock In {\em 2014 IEEE 14th International Working Conference on Source
  Code Analysis and Manipulation}, pages 215--224. IEEE, 2014.

\bibitem{randrianaina2024options}
Georges~Aaron Randrianaina, Djamel~Eddine Khelladi, Olivier Zendra, and Mathieu
  Acher.
\newblock Options matter: Documenting and fixing non-reproducible builds in
  highly-configurable systems.
\newblock In {\em 2024 IEEE/ACM 21st International Conference on Mining
  Software Repositories (MSR)}, pages 654--664. IEEE, 2024.

\bibitem{ren2018automated}
Zhilei Ren, He~Jiang, Jifeng Xuan, and Zijiang Yang.
\newblock Automated localization for unreproducible builds.
\newblock In {\em Proceedings of the 40th International Conference on Software
  Engineering}, pages 71--81, 2018.

\bibitem{ren2022automated}
Zhilei Ren, Shiwei Sun, Jifeng Xuan, Xiaochen Li, Zhide Zhou, and He~Jiang.
\newblock Automated patching for unreproducible builds.
\newblock In {\em Proceedings of the 44th International Conference on Software
  Engineering}, pages 200--211, 2022.

\bibitem{sadowski2018lessons}
Caitlin Sadowski, Edward Aftandilian, Alex Eagle, Liam Miller-Cushon, and Ciera
  Jaspan.
\newblock Lessons from building static analysis tools at google.
\newblock {\em Communications of the ACM}, 61(4):58--66, 2018.

\bibitem{schneider2022experimental}
Jean-Guy Schneider and Sung~Une Lee.
\newblock An experimental comparison of clone detection techniques using java
  bytecode.
\newblock In {\em 2022 29th Asia-Pacific Software Engineering Conference
  (APSEC)}, pages 139--148. IEEE, 2022.

\bibitem{schott2024JNorm}
Stefan Schott, Serena~Elisa Ponta, Wolfram Fischer, Jonas Klauke, and Eric
  Bodden.
\newblock Java bytecode normalization for code similarity analysis.
\newblock 2024.

\bibitem{sharma2025canonicalization}
Aman Sharma, Benoit Baudry, and Martin Monperrus.
\newblock Canonicalization for unreproducible builds in java.
\newblock {\em arXiv preprint arXiv:2504.21679}, 2025.

\bibitem{shi2021experience}
Yong Shi, Mingzhi Wen, Filipe~R Cogo, Boyuan Chen, and Zhen~Ming Jiang.
\newblock An experience report on producing verifiable builds for large-scale
  commercial systems.
\newblock {\em IEEE Transactions on Software Engineering}, 48(9):3361--3377,
  2021.

\bibitem{smaragdakis2010using}
Yannis Smaragdakis and Martin Bravenboer.
\newblock Using datalog for fast and easy program analysis.
\newblock In {\em International Datalog 2.0 Workshop}, pages 245--251.
  Springer, 2010.

\bibitem{vallee2010soot}
Raja Vall{\'e}e-Rai, Phong Co, Etienne Gagnon, Laurie Hendren, Patrick Lam, and
  Vijay Sundaresan.
\newblock Soot: A java bytecode optimization framework.
\newblock In {\em CASCON First Decade High Impact Papers}, pages 214--224.
  2010.

\bibitem{whaley2005using}
John Whaley, Dzintars Avots, Michael Carbin, and Monica~S Lam.
\newblock Using datalog with binary decision diagrams for program analysis.
\newblock In {\em Asian Symposium on Programming Languages and Systems}, pages
  97--118. Springer, 2005.

\bibitem{xiong2022towards}
Jiawen Xiong, Yong Shi, Boyuan Chen, Filipe~R Cogo, and Zhen~Ming Jiang.
\newblock Towards build verifiability for java-based systems.
\newblock In {\em Proceedings of the 44th International Conference on Software
  Engineering: Software Engineering in Practice}, pages 297--306, 2022.

\bibitem{yu2019detecting}
Dongjin Yu, Jiazha Yang, Xin Chen, and Jie Chen.
\newblock Detecting java code clones based on bytecode sequence alignment.
\newblock {\em IEEE Access}, 7:22421--22433, 2019.

\end{thebibliography}

\end{document}